\documentclass[journal]{IEEEtran}

 \usepackage{graphicx} 
  \DeclareGraphicsExtensions{.eps}
  
  \usepackage{subcaption} 
  
  \usepackage{tikz}
  \usepackage{pgfplots,pgfplotstable}
  \usetikzlibrary{pgfplots.groupplots}
  \pgfplotsset{compat=1.5}

\usepackage{multirow}

\usepackage{amsmath}

\usepackage{textcomp}
\usetikzlibrary{shapes,arrows}

\usepackage{empheq} 

\usepackage{stfloats} 

\usepackage[hyphens]{url}
\usepackage[hidelinks]{hyperref}
\hypersetup{colorlinks=true,linkcolor=black,citecolor=black,filecolor=black,urlcolor=black,breaklinks=true}
\usepackage{cite} 

\usepackage{color}

\newcommand{\norm}[1]{\left\lVert#1\right\rVert} 

\usepackage{cancel} 

\usepackage{array} 

\usepackage[export]{adjustbox} 

\usepackage[normalem]{ulem} 

\usepackage{amsfonts} 

\usepackage{diffcoeff}

\usepackage{verbatim}
\usepackage{forest}
\usetikzlibrary{arrows.meta, shapes.geometric, calc, shadows,fit}

\colorlet{mygreen}{green!75!black}
\colorlet{col1in}{red!30}
\colorlet{col1out}{red!40}
\colorlet{col2in}{mygreen!40}
\colorlet{col2out}{mygreen!50}
\colorlet{col3in}{blue!30}
\colorlet{col3out}{blue!40}
\colorlet{col4in}{mygreen!20}
\colorlet{col4out}{mygreen!30}
\colorlet{col5in}{blue!10}
\colorlet{col5out}{blue!20}
\colorlet{col6in}{blue!20}
\colorlet{col6out}{blue!30}
\colorlet{col7out}{orange}
\colorlet{col7in}{orange!50}
\colorlet{col8out}{orange!40}
\colorlet{col8in}{orange!20}
\colorlet{linecol}{blue!60}

\makeatletter
\def\bstctlcite{\@ifnextchar[{\@bstctlcite}{\@bstctlcite[@auxout]}}
\def\@bstctlcite[#1]#2{\@bsphack
 \@for\@citeb:=#2\do{%
   \edef\@citeb{\expandafter\@firstofone\@citeb}%
   \if@filesw\immediate\write\csname #1\endcsname{\string\citation{\@citeb}}\fi}%
 \@esphack}
\makeatother

\begin{document}

\bstctlcite{IEEEexample:BSTcontrol}

\title{Cost Allocation for Inertia and\\   Frequency Response Ancillary Services}

%
\author{Carlos~Matamala,~\IEEEmembership{Student Member,~IEEE},
        Luis~Badesa,~\IEEEmembership{Member,~IEEE},
        Rodrigo~Moreno,~\IEEEmembership{Member,~IEEE},
        and~Goran~Strbac,~\IEEEmembership{Member,~IEEE}
\thanks{C. Matamala and G. Strbac are with the Department of Electrical and Electronic Engineering, Imperial College London, SW7 2AZ London, U.K. (emails: \{c.matamala-vergara, g.strbac\}@imperial.ac.uk). L. Badesa is with the School of Industrial Engineering and Design (\mbox{ETSIDI}), Technical University of Madrid (UPM), 28012 Madrid, Spain. (email: luis.badesa@upm.es). R. Moreno is with the Electrical Engineering Department at the University of Chile and Instituto Sistemas Complejos de Ingeniería (ISCI), Santiago, Chile. (email: rmorenovieyra@ing.uchile.cl).}
\thanks{C. Matamala was supported by the National Agency for Research and Development (ANID) through scholarship ANID/PROGRAMA BECAS CHILE DOCTORADO under Grant 2020-72210414. L. Badesa was supported by the Horizon Europe project “TRANSIT - TRANSITion to sustainable future through training and education” (grant no. 101075747). R. Moreno was supported by the Instituto de Sistemas Complejos de Ingenieria [FIN ANID AFB 230002] and FONDECYT 1231924. G. Strbac acknowledges support from two major research projects (a) Integrated Development of Low-Carbon Energy Systems (EP/R045518/1) and (b) UK Energy Research Centre Phase 4 (EP/S029575/1).}
        
%
\vspace{-8mm}
}

%
%

\markboth{IEEE Transactions on Energy Markets, Policy and Regulation, January~2024}%
{Shell \MakeLowercase{\textit{et al.}}: Bare Demo of IEEEtran.cls for IEEE Journals}
%



\maketitle
\begin{abstract}
The reduction in system inertia is creating an important market for frequency-containment Ancillary Services (AS) such as enhanced frequency response (e.g.,~provided by battery storage), traditional primary frequency response and inertia itself. This market presents an important difference with the energy-only market: while the need for energy production is driven by the demand from consumers, frequency-containment AS are procured because of the need to deal with the largest generation/demand loss in the system (or smaller losses that could potentially compromise frequency stability). Thus, a question that arises is: who should pay for frequency-containment AS? In this work, we 
propose a cost-allocation methodology based on the nucleolus concept, in order to distribute the total payments for frequency-containment AS among all generators or loads that create the need for these services. 
It is shown that this method complies with necessary properties for the AS market, such as avoidance of cross-subsidies and maintaining players in this cooperative game. Finally, we demonstrate its practical applicability through a case study for the Great Britain power system, while comparing its performance with two alternative mechanisms, namely proportional and Shapley value cost allocation. 
\end{abstract}

\vspace*{-1mm}

%
\IEEEpeerreviewmaketitle

\vspace*{-3mm}
\section*{Nomenclature}
\addcontentsline{toc}{section}{Nomenclature}

\vspace*{-3mm}
\subsection*{Indices and Sets}
\begin{IEEEdescription}[\IEEEusemathlabelsep\IEEEsetlabelwidth{$P_{s}^\textrm{cha}, P_{s}^\textrm{dis}$}]
    \item[$g, \mathcal{G}$] Index and set of generators.
    \item[$r, \mathcal{R}$] Index and set of RES.
    \item[$s, \mathcal{S}$] Index and set of storage.
    \item[$i,\mathcal{M}_t,\mathcal{N}_t$] Index, subset, and set of total dispatched units at hour $t$.
    \item[$t, \mathcal{T}, \textrm{t}^\textrm{end}$] Index, set of time horizon, and final time.
\end{IEEEdescription}

\vspace*{-5mm}
\subsection*{Parameters}
\begin{IEEEdescription}[\IEEEusemathlabelsep\IEEEsetlabelwidth{$P_{s}^\textrm{cha}, P_{s}^\textrm{dis}$}]
    \item[$\Delta f_{\textrm{max}}$] Maximum admissible frequency deviation at the nadir (Hz).
    \item[$\mathrm{\eta_{s}^{cha}}, \mathrm{\eta_{s}^{dis}}$] Charge and discharge efficiency for storage s.
    \item[$\lambda^\textrm{E}_g, \lambda^\textrm{E}_r, \lambda^\textrm{E}_s$] Energy price offer of generator $g$, RES $r$, and storage $s$ (\pounds/MWh). 
    \item[$\lambda^\textrm{H}_g, \lambda^\textrm{H}_s$] Inertia price offer of generator $g$, and storage $s$ (\pounds/MWs).
     \item[$\lambda^\textrm{PFR}_g, \lambda^\textrm{PFR}_s, \lambda^\textrm{EFR}_s$] \qquad FR price offer of generator $g$, and storage $s$ (\pounds/MW).
    \item[$C(\cdot)$] Generic cost function.
    \item[$\textrm{CF}_{r,t}$] Capacity factor at hour $t$ for RES $r$.
    \item[$\textrm{EFR}^\textrm{max}_s$] EFR capacity of storage $s$ (MW).
    \item[$\mathrm{E}_{s}^\mathrm{max}, \mathrm{E}_{s}^\mathrm{min}$] Max/min charge level for storage $s$ (MWh).
    \vspace{0.5mm}
    \item[$f_0$] Nominal power grid frequency (Hz).
    \item[$\textrm{H}_g, \textrm{H}_s$] Inertia constant of generator $g$ and storage $s$ (s).
    \vspace{0.5mm}
    \item[$\textrm{P}^\textrm{D}_t$] Total system demand at hour $t$ (MW).
    \item[$\textrm{PFR}^\textrm{max}_g, \textrm{PFR}^\textrm{max}_s$] \qquad PFR capacity of generator $g$ and storage $s$ (MW).
    \vspace{0.5mm}
    \item[$\textrm{P}^\textrm{max}_g, \textrm{P}^\textrm{max}_r, \textrm{P}^\textrm{max}_s$] \qquad Rated power of generator $g$, RES $r$, and storage $s$ (MW).
    \item[$\textrm{P}^\textrm{msg}_g,  \textrm{P}_{s}^\textrm{msg}$] Minimum stable generation of generator $g$, and storage $s$ (MW).
    \vspace{0.5mm}
    \item[$\textrm{RoCoF}_\textrm{max}$]  Maximum admissible RoCoF (Hz/s).
    \item[$\textrm{T}_\textrm{EFR}, \textrm{T}_\textrm{PFR}$]  Delivery time of EFR and PFR (s).
    \vspace{0.5mm}
    \item[$\textrm{T}_g^\textrm{mdt}, \textrm{T}_g^\textrm{mut}, \textrm{T}_g^\textrm{st}$] \quad Minimum down, minimum up, and start-up time for generator $g$ (h).
\end{IEEEdescription}

\vspace*{-5mm}
\subsection*{Variables \normalfont{(continuous and time-dependent unless otherwise indicated)}} 
\begin{IEEEdescription}[\IEEEusemathlabelsep\IEEEsetlabelwidth{$P_{s}^\textrm{cha}, P_{s}^\textrm{dis}$}]
    \item[$\lambda^{E}_t$]  Market-clearing energy price (£/MWh).
    \item[$\lambda^{EFR}_t, \lambda^{PFR}_t$] \quad Market-clearing EFR and PFR price (£/MW).
    \item[$\lambda_t^{\textrm{H}}$] Market-clearing inertia price (£/MW$\textrm{s}$).
    \item[$\mu_{t}^\textrm{nadir}$] Dual variables for the SOC nadir constraint.
    \item[$\mu^\textrm{q-s-s}_t$] Dual variable for the q-s-s constraint.
    \item[$\mu^\textrm{RoCoF}_t$] Dual variable for the RoCoF constraint.
    \item[$\varphi_{i,t}^{\textrm{Pro}}, \varphi_{i,t}^{\textrm{SV}}, \varphi_{i,t}^{\textrm{Nucl}}$] \qquad \quad Proportional, Shapley value, and nucleolus cost allocation for unit $i$ (£).     
    \item[$\omega^\textrm{Loss}_t$] Dual variable associated with the incremental change in the AS market.
    \item[$E_{s,t}$] State of charge of storage $s$ (MWh).
    \vspace{0.5mm}
    \item[$P_{g,t}, P_{r,t}$] Power produced by generator $g$ and RES $r$ (MW).
    \item[$P_{s,t}^\textrm{cha}, P_{s,t}^\textrm{dis}$]  Charge and discharge power of storage $s$ (MW).
    \item[$P^\textrm{Loss}_t$] Largest power infeed (MW).
    \item[$PFR_{g,t}, PFR_{s,t}, EFR_{s,t}$] \qquad \qquad \qquad Headroom in generator $g$ and storage $s$ for providing FR (MW).
    \vspace{0.5mm}
    \item[$y_{g,t}, y_{g,t}^\textrm{sd}, y_{g,t}^\textrm{sg}, y_{g,t}^\textrm{st}$] \qquad \quad Binary variables. Commitment, shut-down, start generating, and start-up state of generator $g$.
    \item[$y_{s,t}^\textrm{cha}, y_{s,t}^\textrm{dis}$]  Binary variables. Charging and discharging state of storage $s$.
\end{IEEEdescription}

\vspace*{-5mm}
\subsection*{Linear Expressions \normalfont{(linear combinations of decision variables)}}
\begin{IEEEdescription}[\IEEEusemathlabelsep\IEEEsetlabelwidth{$P_{s}^\textrm{cha}, P_{s}^\textrm{dis}$}]
    \item[$EFR_t, PFR_t$]  \qquad \quad Aggregate headroom for providing EFR and PFR (MW).
    \vspace{0.5mm}
    \item[$H_t$]  Aggregate synchronous inertia (MW$\textrm{s}$). 
\end{IEEEdescription}

\section{Introduction}
%
%
%
%


\IEEEPARstart{B}{alancing} generation and demand is one of the main challenges that system operators are facing in power systems with high penetration of Renewable Energy Sources (RES) \cite{SCUC_IEEETaskForce,mohandes2019review, zhongming2017roadmap}. Frequency must always be close to the nominal operating point to avoid customer disruptions if a generation or load outage occurs. Given that most RES, such as wind and solar photovoltaics do not provide inertia, power systems are becoming more vulnerable to frequency deviations, making frequency-containment Ancillary Services (AS) more valuable during hours with high RES penetration. The Australian system cascading in 2018 and the situation in Great Britain (GB) during the COVID-19 national lockdown are good examples of this trend \cite{LuisCovid,MancarellaFragile}.

Electricity markets have failed to provide clear incentives to increase frequency security, and current market designs are unsuitable for players to invest in AS provision \cite{billimoria2020market}. Moreover, consideration of services such as inertia, which strictly depends on the commitment of generators, challenge current pricing mechanisms, as it is necessary to deal with non-convexities that could create the need for uplifts or side-payments \cite{hogan2003minimum, ONeillNonconvex}.

In this vein, pricing frequency-containment AS is a topic recently studied in the literature. In \cite{LuisPricing}, the authors developed a Mixed-Integer Second-Order Cone (MISOC) program, which, after relaxing the binary variables, represents a convex problem. Prices for inertia and frequency response (FR) with different delays and speeds were computed through dual variables. In \cite{LuisSyntPricing}, an improved version of the former paper was developed, considering participation of RES in the AS market, providing services such as FR and synthetic inertia. This work shows that, for linear cost functions and these convex frequency-security constraints, computing prices through relaxing the MISOC problem is equivalent to obtaining convex hull prices, i.e., the problem that minimises uplift payments \cite{gribik2007market,hua2016convex}. Furthermore, a pricing methodology for inertia that considers a chance-constrained stochastic unit commitment was proposed in \cite{LiangInertiaPrice}.

All these works assume that the AS market is eventually paid entirely by electricity consumers, even though the need for AS is not triggered by marginal/incremental increases or decreases of electricity demand. Instead, under-frequency containment AS are triggered by generator outages. One of the first discussions regarding cost allocation of contingency reserve against generator outages was done in~\cite{GoranReserve}. This cost allocation is based on the generating units' capacity and unavailability. The allocation mechanism considers a proportional expected cost allocation, as it is based on the expected undelivered power generation as the main driver for the mechanism. A similar idea was presented in \cite{hirst2003allocating}, in which the authors propose to allocate costs considering two components: (i) the number and size of outages in the last year, considering a proportional expected cost allocation; and (ii) considering the size of the dispatch at each hour, for which both proportional and sequential cost allocation are considered. 

Even though these works recognise the trigger of the need for AS, and propose allocation mechanisms for AS costs based on the potential `harm' that different players do to the system, they do not focus on the desired attributes of a cost allocation mechanism \cite{moulin1991axioms}. Attributes such as \textit{equity}, which considers that no subgroup of market players subsidises any other, or \textit{consistency}, which means that no subgroup has any incentive to leave the grand coalition, are in fact relevant for this task~\cite{hu2006allocation}. Therefore, more sophisticated allocation methodologies such as Shapley value and nucleolus are worth being considered, given that these methods seek these desirable properties \cite{young1985cost}.

Given this context, the present article provides the first rigorous formulation and analysis of three main cost allocation methodologies applied to the frequency-containment AS market. First, we consider the formulation introduced in \cite{LuisPricing, LuisSyntPricing} to calculate the size of the AS market (i.e., the total cost) that each generator's outage would create. Even though system operators only focus on the single largest plausible loss (given that AS markets are typically defined by the $N$-1 security standard), a cost allocation mechanism must consider that this level of security also covers the occurrence of the second largest loss, and smaller losses henceforth. Through three different cost allocation methodologies, namely proportional, Shapley value and nucleolus, we compute the distribution of these costs among all generators. At the same time, we demonstrate that the proposed market framework provides incentives for market participants to supply services such as inertia, Enhanced Frequency Response (EFR) and Primary Frequency Response (PFR). 

It is worth clarifying that, while this work focuses on under-frequency containment AS, and therefore on cost allocation for generators, the discussion would be equivalent for cost allocation among loads. Frequency drops are the main concern in most grids, but over-frequency events are modelled in a symmetrical manner, simply considering a sudden load disconnection instead of a generation outage. The same principles and models would apply to the over-frequency AS case.

The main contributions of this work are:

\begin{enumerate}
    \item To propose a nucleolus-based methodology for distributing the total AS cost among generators, which puts in place appropriate incentives for market participants to inflict as little `harm' as possible to the grid in the form of potentially large generation outages.
    \item To provide theoretical guarantees that this cost allocation mechanism provides necessary properties for the AS market, including avoidance of cross-subsidies while maintaining players in this cooperative game.
    \item To showcase the applicability of the proposed cost allocation through a relevant case study of the future British grid, demonstrating that system-integration costs (the cost of integrating a unit into a power grid) of large plants can be accounted for within the AS market.
\end{enumerate}

The remainder of this paper is structured as follows: Section~\ref{sec:Methodology} describes the mathematical methodology that defines the AS market design. The cost allocation mechanisms under consideration are discussed in Section~\ref{sec:Mechanisms}. Section~\ref{sec:Results} illustrates the applicability of the AS market design and cost allocation through a relevant case study representative of the GB power system. Finally, Section~\ref{sec:Conclusion} gives the conclusion.

\section{Definition of Ancillary Services Market} \label{sec:Methodology}

In this section, we present a multi-period frequency-secured Unit Commitment (UC) with hourly resolution, formulated as an MISOC problem due to the frequency-security constraints. This generic model, which may consider any type of generation or storage technologies and therefore is applicable to any power grid, will be used to compute the size of the AS market for each possible contingency. 

Three different frequency-containment AS are considered here. Inertia, which refers to the kinetic energy stored in synchronous machines and released in the event of a frequency drop, supports the balance between generation and demand at all times \cite{KundurBook}. We distinguish two different FR services: a fast service provided by technologies based on power electronics, referred to as EFR, and the traditional service provided by synchronous generators, PFR. 

Regarding the mathematical notation, variables in parenthesis represent dual variables associated with a certain constraint. We use the term \textit{system-wide constraints} for constraints related to the system requirements, such as respecting frequency limits or balancing load and generation, while \textit{private constraints} are associated directly to specific market players.

\subsection{Frequency-secured UC formulation} \label{sec:FreqSecUC}
The objective function considered for the UC is the following linear function:
\begin{alignat}{3}
    &  {\text{min}} && \Biggl\{ \sum_{t \in \mathcal{T}}  \Biggr[
     \sum_{g \in \mathcal{G}} \biggl( \lambda^\textrm{E}_g\cdot P_{g,t}  +  \lambda^\textrm{H}_g \cdot \textrm{P}^\textrm{max}_g \cdot \textrm{H}_g  \cdot y_{g,t}  \nonumber \\[3pt]
    & &&+ \lambda^\textrm{PFR}_g \cdot PFR_{g,t} \biggl) + \sum_{r \in \mathcal{R}}  \lambda^\textrm{E}_r\cdot P_{r,t}   + \sum_{s \in \mathcal{S}} \biggl(  \lambda^\textrm{E}_s\cdot P_{s,t}^\textrm{dis}  \nonumber \\[3pt] 
    & && +  \lambda^\textrm{H}_s \cdot \textrm{P}^\textrm{max}_s \cdot \textrm{H}_s \cdot (y_{s,t}^\textrm{cha} + y_{s,t}^\textrm{dis})  
    +  \lambda^\textrm{PFR}_s \cdot PFR_{s,t} \nonumber \\[3pt] 
    & && + \lambda^\textrm{EFR}_s \cdot EFR_{s,t} \biggl)  \Biggr]  \Biggl\}
    \label{eq:objFunc}
\end{alignat}

The energy balance system-wide constraint is defined as:
\begin{alignat}{3}
    & \sum_{g \in \mathcal{G}}P_{g,t} + \sum_{r \in \mathcal{R}} P_{r,t} + \sum_{s \in \mathcal{S}} \left( P_{s,t}^\textrm{dis} -  P_{s,t}^\textrm{cha} \right)= \textrm{P}^\textrm{D}_t:  (\lambda_{t}^{\textrm{E}}) &&  \;\forall t  \label{eq:e_balance}
\end{alignat}

Thermal generators are defined by the private constraints~(\ref{eq:CommitmentTime})-(\ref{eq:Binary_y}), which apply $\forall g \in \mathcal{G}, t \in  \mathcal{T}$. Constraint~(\ref{eq:CommitmentTime}) establishes the relation between online, start generating and shut down units at each hour. Constraint~(\ref{eq:StartUpIndicator}) represents that a unit starts generating if it was started up $\textrm{T}_g^\textrm{st}$ hours before.  Constraint~(\ref{eq:StartUp}) enforces the minimum down time, which means that a unit is allowed to be started up if it was ‘off’ in the previous time-step and has been ‘off’ for at least $\textrm{T}_g^\textrm{mdt}$ hours. Constraint~(\ref{eq:DownTime}) defines the minimum up time, meaning that a unit is allowed to be shut down if it was generating in the previous time-step, but also has been generating for at least $\textrm{T}_g^\textrm{mut}$ hours. Constraint~(\ref{eq:gen}) implies that, if the unit is committed, its generation lies between its minimum stable generation and maximum capacity. Finally, constraints (\ref{eq:pfr1_gen}) and (\ref{eq:pfr2_gen}) represent the maximum PFR deliverable and the available margin to provide it, respectively. 

RES generation is modelled with an hourly generation profile defined by the capacity factor $\textrm{CF}_{r,t}$ as shown in private constraint~(\ref{eq:res_gen}), which applies $\forall r \in \mathcal{R}, t \in  \mathcal{T}$. For simplicity, in the case study in Section~\ref{sec:Results}, we consider a single RES profile for each of the three RES technologies: offshore wind, onshore wind, and solar photovoltaic.

Energy storage units are defined by private constraints~(\ref{eq:e_stor_minmax})-(\ref{eq:e_stor_end}). Constraints~(\ref{eq:e_stor_minmax})-(\ref{eq:Binary_ys}) apply $\forall s \in \mathcal{S}, t \in  \mathcal{T}$, while (\ref{eq:e_stor_ini})-(\ref{eq:e_stor_end}) apply $\forall s \in \mathcal{S}$. Constraint~(\ref{eq:e_stor_minmax}) states that the energy stored should lie between the minimum and the maximum of the storage unit. The variation of state of charge due to the operation of the storage unit over time is modelled through constraint~(\ref{eq:e_stor}). Charge and discharge limits are modelled with constraints (\ref{eq:genc}) and (\ref{eq:gend}), respectively. Constraints~(\ref{eq:Sum_Binary_ys})-(\ref{eq:Binary_ys}) state that the storage unit can be either charging or discharging, but not in both states simultaneously. Constraints (\ref{eq:e_stor_ini}) and (\ref{eq:e_stor_end}) enforce the initial and final state of charge, respectively. For pumped hydro energy storage (PHES) units, the maximum and available margin for PFR are defined by constraints (\ref{eq:pfr1_stor}) and (\ref{eq:pfr2_stor}), respectively. Constraint~(\ref{eq:pfr2_stor}) implies that there is available PFR either when the PHES unit is discharging, or when the unit is charging, in which case the available PFR will be limited by the charging power in that hour. In the case of Battery Energy Storage Systems (BESS), maximum EFR and available margin are determined by constraints (\ref{eq:efr1_stor}) and (\ref{eq:efr2_stor}), respectively. Note that constraint~(\ref{eq:efr2_stor}) accounts for the fact that, as BESS are based on power electronics, these units can swiftly switch from charging to discharging, therefore providing a higher volume of EFR if operating in charging mode. 

\vspace*{-30mm}

\begin{alignat}{2}
    & y_{g,t} = y_{g,t-1} + y_{g,t}^\textrm{sg} - y_{g,t}^\textrm{sd} && \; \;\forall g,t\label{eq:CommitmentTime}\\[3pt]
    & y_{g,t}^\textrm{sg} = y_{g,t-\textrm{T}_g^\textrm{st}}^\textrm{st} &&\; \;\forall g,t \label{eq:StartUpIndicator}\\
    & y_{g,t}^\textrm{st} \leq
    \left[1-y_{g,t-1}\right] - \hspace{-1mm} \sum_{j=t-\textrm{T}_g^\textrm{mdt}}^t y_{g,j}^\textrm{sd}: (\psi^{\textrm{mdt}}_{g,t}) &&  \; \;\forall g,t\label{eq:StartUp}\\ 
    & y_{g,t}^\textrm{sd} \leq 
    y_{g,t-1} - \hspace{-1mm} \sum_{j=t-\textrm{T}_g^\textrm{mut}}^{t}y_{g,j}^\textrm{sg} &&  \; \;\forall g,t\label{eq:DownTime}\\[3pt]
    & y_{g,t} \cdot \mathrm{P}_{g}^\mathrm{msg}  \leq P_{g,t} \leq y_{g,t} \cdot \mathrm{P}_{g}^\mathrm{max} &&  \; \;\forall g,t  \label{eq:gen}\\[3pt]
    & PFR_{g,t} \leq y_{g,t} \cdot \mathrm{PFR}_{g}^\mathrm{max} &&  \; \; \forall g,t  \label{eq:pfr1_gen}  \\[3pt]
    & PFR_{g,t} \leq y_{g,t} \cdot \mathrm{P}_{g}^\mathrm{max} - P_{g,t} &&  \; \; \forall g,t   \label{eq:pfr2_gen}\\[3pt]
    & y_{g,t},\; y_{g,t}^\textrm{st},\; y_{g,t}^\textrm{sg},\; y_{g,t}^\textrm{sd} \in  \{0, 1\} && \; \;\forall g,t \label{eq:Binary_y}\\[3pt]
    & P_{r,t} \leq \textrm{CF}_{r,t} \cdot \mathrm{P}_{r}^\mathrm{max}: (\psi^{\textrm{CF}}_{r,t}) && \; \; \forall r,t  \label{eq:res_gen}
\end{alignat}

\begin{alignat}{2}
    & \mathrm{E}_{s}^\mathrm{min} \leq E_{s,t} \leq \mathrm{E}_{s}^\mathrm{max}: \left( \psi^{\textrm{min}(E_s)}_{s,t}, \psi^{\textrm{max}(E_s)}_{s,t} \right) && \; \;\forall s,t \label{eq:e_stor_minmax}\\[3pt]
    & E_{s,t}  = E_{s,t-1} + P_{s,t}^\textrm{cha} \cdot \mathrm{\eta}_{s}^\textrm{cha} - P_{s,t}^\textrm{dis}/\mathrm{\eta}_{s}^\textrm{dis}  && \; \;\forall s,t \label{eq:e_stor}\\[3pt]
    & y_{s,t}^\textrm{cha} \cdot \mathrm{P}_{s}^\mathrm{msg}  \leq P_{s,t}^\textrm{cha} \leq y_{s,t}^\textrm{cha} \cdot\mathrm{P}_{s}^\mathrm{max} && \; \;\forall s,t \label{eq:genc} \\[3pt] 
    & y_{s,t}^\textrm{dis} \cdot \mathrm{P}_{s}^\mathrm{msg}  \leq P_{s,t}^\textrm{dis} \leq y_{s,t}^\textrm{dis} \cdot \mathrm{P}_{s}^\mathrm{max} && \; \;\forall s,t \label{eq:gend}\\[3pt] %
    & PFR_{s,t} \leq (y_{s,t}^\textrm{dis}+y_{s,t}^\textrm{cha}) \cdot \mathrm{PFR}_{s}^\mathrm{max}  && \; \;\forall s,t \label{eq:pfr1_stor} \\[3pt]
    & PFR_{s,t} \leq y_{s,t}^\textrm{dis} \cdot \mathrm{P}_{s}^\mathrm{max} -  P_{s,t}^\textrm{dis} +  P_{s,t}^\textrm{cha}  && \; \;\forall s,t \label{eq:pfr2_stor} \\[3pt]
    & EFR_{s,t} \leq  (y_{s,t}^\textrm{cha} + y_{s,t}^\textrm{dis})  \cdot \mathrm{EFR}_{s}^\mathrm{max} \label{eq:efr1_stor} && \; \;\forall s,t \\[3pt]
    & EFR_{s,t}\leq( y_{s,t}^\textrm{dis}+y_{s,t}^\textrm{cha}) \cdot \mathrm{P}_{s}^\mathrm{max}-P_{s,t}^\textrm{dis}+P_{s,t}^\textrm{cha} && \; \;\forall s,t \label{eq:efr2_stor}\\[3pt]
    & y_{s,t}^\textrm{cha} + y_{s,t}^\textrm{dis} \leq 1 : (\psi^{\textrm{(dis-cha)}}_{s,t} ) && \; \;\forall s,t \label{eq:Sum_Binary_ys}\\[3pt]
    & y_{s,t}^\textrm{cha},\; y_{s,t}^\textrm{dis} \in  \{0, 1\} && \; \;\forall s,t \label{eq:Binary_ys} \\[3pt]
    & E_{s,1}  \leq \mathrm{E}_{s}^\textrm{ini}: \left( \psi^{\textrm{ini}}_{s,1} \right) && \; \;\forall s  \label{eq:e_stor_ini}\\
    & E_{s,\textrm{t}^\textrm{end}}  \geq \mathrm{E}_{s}^\textrm{end}: \left( \psi^{\textrm{end}}_{s,\textrm{t}^\textrm{end}} \right) && \; \;\forall s \label{eq:e_stor_end}
\end{alignat}

\subsection{Ancillary services for frequency containment} \label{sec:TotalProvisionAS}
Synchronous inertia is proportional to a thermal unit's inertia constant and rated power, and depends on its commitment state. PHES units can provide inertia in any state, whether they are discharging or charging. Total system inertia is defined by the following system-wide constraint: 

\begin{alignat}{3}
    & H_t = &&  \sum_{ g \in \mathcal{G} } \textrm{H}_g \cdot \textrm{P}^\textrm{max}_g \cdot y_{g,t} && \nonumber\\
    &  && + \sum_{ s \in \mathcal{S} } \textrm{H}_s \cdot \textrm{P}^\textrm{max}_s \cdot (y_{s,t}^\textrm{dis} + y_{s,t}^\textrm{cha}):  (\lambda_{t}^{H}) && \quad \forall t \label{eq:SyncInertia}
\end{alignat}

PFR, provided by synchronous generators and PHES, is defined by system-wide constraint~(\ref{eq:pfr_total}), while EFR, provided by BESS, is defined by system-wide constraint~(\ref{eq:efr_total}):

\begin{alignat}{3}
    &PFR_t = &&  \sum_{ g \in \mathcal{G} } PFR_{g,t} + \sum_{ s \in \mathcal{S} } PFR_{s,t}:  (\lambda_{t}^{PFR})  && \quad \forall t \label{eq:pfr_total}\\[3pt]
    &EFR_t = &&  \sum_{ s \in \mathcal{S} } EFR_{s,t}:  (\lambda_{t}^{EFR})  && \quad \forall t \label{eq:efr_total}
\end{alignat}

\subsection{Frequency-security constraints} \label{sec:FreqConstraints}
The set of system-wide constraints that determine the requirement of AS for frequency containment, and couple the different services, is discussed here.

First, decision variable $P^\textrm{Loss}_{t}$, which determines the largest possible contingency, is limited by the maximum possible loss at each hour, e.g., $\textrm{P}_t^\textrm{Loss}=1.8\textrm{GW}$ in GB. The proposed market design establishes an inequality constraint to lead to non-negative dual variables associated with this constraint:
\begin{equation} \label{eq:maxPloss}
    \textrm{P}_t^\textrm{Loss}
    \leq
     P^\textrm{Loss}_{t} :  \; \; (\omega^\textrm{Loss}_t) \; \;\forall t
\end{equation}

At the very instant of the outage, the Rate-of-Change-of-Frequency (RoCoF) constraint guarantees that no islanding-protection scheme will be triggered, as this could exacerbate the frequency excursion \cite{LuisSyntPricing}. This constraint is given by: 
\begin{equation} \label{eq:Rocof}
    H_t 
    \geq
    \frac{ P^\textrm{Loss}_{t} \cdot f_0}{2\cdot\textrm{RoCoF}_\textrm{max}  }:  \; \; (\mu^\textrm{RoCoF}_t) \; \;\forall t
\end{equation}

The nadir constraint avoids the activation of under-frequency load shedding and is formulated in~(\ref{eq:nadirSOC}) as a standard SOC. This constraint was first introduced in \cite{LuisMultiFR}. 

\begin{multline} \label{eq:nadirSOC}
    \norm{
    \begin{bmatrix} 
        \frac{1}{f_0}  
        & \frac{-\textrm{T}_\textrm{EFR}}{4\Delta f_\textrm{max}}  
        &  \frac{-1}{\textrm{T}_\textrm{PFR}} 
        & 0 \\[5pt]
        0 &\frac{-1}{\sqrt{\Delta f_\textrm{max}}} & 0 &\frac{1}{\sqrt{\Delta f_\textrm{max}}}
    \end{bmatrix}
    \begin{bmatrix} 
        H_t   
        \\ EFR_t 
        \\ PFR_t 
        \\  P^\textrm{Loss}_{t} 
    \end{bmatrix} 
    } 
    \\
    \leq  \begin{bmatrix} 
        \frac{1}{f_0}  
        & \frac{-\textrm{T}_\textrm{EFR}}{4\Delta f_\textrm{max}}  
        & \frac{1}{\textrm{T}_\textrm{PFR}} 
        & 0 
    \end{bmatrix}
    \begin{bmatrix} 
        H_t    
        \\ EFR_t 
        \\ PFR_t 
        \\  P^\textrm{Loss}_{t} 
    \end{bmatrix}   
    \\
    : \; \;(\mu_{1,t}^\textrm{nadir}, \;\mu_{2,t}^\textrm{nadir}, \;\mu_{3,t}^\textrm{nadir}) \; \; \forall t
\end{multline}
On a broader time window, a power equilibrium must be reached, for which total FR must be, at least, equal to the size of the outage. Frequency would then stabilise at a value typically lower than nominal, thus the frequency quasi-steady-state (q-s-s) constraint is:

\begin{equation} \label{eq:qss}
    EFR_t + PFR_t 
    \geq
     P^\textrm{Loss}_{t}: \; \; (\mu^\textrm{q-s-s}_t) \; \; \forall t
\end{equation}

Finally, it is important to clarify that if a different set of frequency-security constraints was used, the cost allocation methodologies proposed in Section~\ref{sec:Mechanisms} would still apply. The formulation presented in this Section~\ref{sec:FreqConstraints} is however particularly convenient to compute AS prices, as discussed next.

\subsection{Prices for inertia and frequency response}
The methodology in \cite{LuisPricing} is used here to compute AS prices. To do this, we relax binary variables (\ref{eq:Binary_y}) and (\ref{eq:Binary_ys}), which are re-stated in (\ref{eq:Continuous_y})-(\ref{eq:Continuous_y_sd}) for generators and (\ref{eq:Continuous_ys_cha})-(\ref{eq:Continuous_ys_dis}) for storage, then obtaining a convex formulation.

\begin{alignat}{2}
    & 0 \leq y_{g,t} \leq 1:    \; \; \left(\psi^{\textrm{min}(y_g)}_{g,t}, \psi^{\textrm{max} (y_g)}_{g,t} \right) && \; \; \forall g,t \label{eq:Continuous_y} \\
    & 0 \leq y_{g,t}^\textrm{st} \leq 1:    \; \; \left(\psi^{\textrm{min}(y^\textrm{st}_g)}_{g,t}, \psi^{\textrm{max} (y^\textrm{st}_g)}_{g,t} \right) && \; \; \forall g,t \label{eq:Continuous_y_st} \\
    & 0 \leq y_{g,t}^\textrm{sg} \leq 1:    \; \; \left(\psi^{\textrm{min}(y^\textrm{sg}_g)}_{g,t}, \psi^{\textrm{max} (y^\textrm{sg}_g)}_{g,t} \right) && \; \; \forall g,t \label{eq:Continuous_y_sg} \\
    & 0 \leq y_{g,t}^\textrm{sd} \leq 1:    \; \; \left(\psi^{\textrm{min}(y^\textrm{sd}_g)}_{g,t}, \psi^{\textrm{max} (y^\textrm{sd}_g)}_{g,t} \right) && \; \; \forall g,t \label{eq:Continuous_y_sd} \\
    & 0 \leq y_{s,t}^\textrm{cha}  \leq 1:   \; \; \left(\psi^{\textrm{min}({y^\textrm{cha}_s)}}_{s,t}, \psi^{\textrm{max}( y^\textrm{cha}_s)}_{s,t} \right)  && \; \; \forall s,t \label{eq:Continuous_ys_cha} \\
    & 0 \leq y_{s,t}^\textrm{dis}  \leq 1:   \; \; \left(\psi^{\textrm{min}(y^\textrm{dis}_s)}_{s,t}, \psi^{\textrm{max}(y^\textrm{dis}_s)}_{s,t} \right) && \; \; \forall s,t \label{eq:Continuous_ys_dis}
\end{alignat}

To derive AS prices from eqs.~(\ref{eq:SyncH_price})-(\ref{eq:priceEFR}), we consider the Karush-Kuhn-Tucker stationarity conditions of the Lagrangian function, which is included in eq.~(\ref{eq:Lagrangian}) within Appendix~\ref{ap:Lagrangian_systemwide}.

\begin{alignat}{4}
    & \diffp{\mathcal{L}}{H_t} &&= 0  && \Rightarrow \lambda_{t}^{H}  &&=  \frac{\mu_{3,t}^\textrm{nadir} -\mu_{1,t}^\textrm{nadir}}{f_0} +  \mu^\textrm{RoCoF}_t \; \;\forall t \label{eq:SyncH_price} \\
    & \diffp{\mathcal{L}}{PFR_t} &&= 0  && \Rightarrow \lambda_{t}^{PFR} &&=  \frac{\mu_{3,t}^\textrm{nadir} + \mu_{1,t}^\textrm{nadir}}{\textrm{T}_\textrm{PFR}} + \mu^\textrm{q-s-s}_t \; \;\forall t \label{eq:PFR_price} \\ 
    &\diffp{\mathcal{L}}{EFR_t} &&= 0      && \Rightarrow  \lambda_{t}^{EFR}  &&= \frac{\left( \mu_{1,t}^\textrm{nadir} - \mu_{3,t}^\textrm{nadir} \right) \cdot \textrm{T}_\textrm{EFR}}{4\Delta f_\textrm{max}}  \nonumber\\
    & && && &&  + \frac{\mu_{2,t}^\textrm{nadir}}{\sqrt{\Delta f_\textrm{max}}}  + \mu^\textrm{q-s-s}_t  \; \quad\forall t \label{eq:priceEFR} \\
    &\diffp{\mathcal{L}}{P^\textrm{Loss}_{t}} &&= 0  && \Rightarrow \omega^\textrm{Loss}_t  &&= \frac{\mu^\textrm{RoCoF}_t \cdot f_0}{2\cdot\textrm{RoCoF}_\textrm{max}} + \frac{ \mu_{2,t}^\textrm{nadir} }{\sqrt{\Delta f_\textrm{max}}} \nonumber\\
    & && && &&+ \mu^\textrm{q-s-s}_t  \; \quad\forall t \label{eq:price_loss}
\end{alignat}

\subsection{Revenues from frequency-containment services}
The relaxation of the problem shown in \ref{sec:FreqSecUC}, \ref{sec:TotalProvisionAS}, and \ref{sec:FreqConstraints} with  (\ref{eq:Continuous_y})-(\ref{eq:Continuous_ys_dis}) enables the use of the strong duality condition \cite{andersen2002notes}. Reordering the equality relation for strong duality, we obtain the expression in eq.~(\ref{eq:strong_dual}). As can be seen, energy and AS payments can cover, in this convex formulation, the system costs and the profits for each type of technology, considering thermal, RES, and storage technologies. 

\begin{multline} \label{eq:strong_dual}
    \left. 
    \sum_{t \in  \mathcal{T}} \left( \textrm{P}^\textrm{D}_t \cdot \lambda_{t}^{\textrm{E}}  +
    \textrm{P}_t^\textrm{Loss} \cdot \omega^\textrm{Loss}_t \right) 
    \hspace*{30mm}\right\}{\parbox{1.5cm}{\scriptsize {energy and \\AS payments}}}\\
    \left. 
    \begin{array}{cc}
    \displaystyle   = \sum_{t \in \mathcal{T}}  \sum_{g \in \mathcal{G}} \biggl( \lambda^\textrm{E}_g\cdot P_{g,t}  +  \lambda^\textrm{H}_g \cdot \textrm{P}^\textrm{max}_g \cdot \textrm{H}_g  \cdot y_{g,t} \\ [\jot] 
    \displaystyle + \lambda^\textrm{PFR}_g \cdot PFR_{g,t} \biggl) + \sum_{t \in \mathcal{T}} \sum_{r \in \mathcal{R}}  \lambda^\textrm{E}_r\cdot P_{r,t} \\[\jot] 
    \displaystyle \sum_{t \in \mathcal{T}} \sum_{s \in \mathcal{S}} \biggl(  \lambda^\textrm{E}_s\cdot P_{s,t}^\textrm{dis} 
     +  \lambda^\textrm{H}_s \cdot \textrm{P}^\textrm{max}_s \cdot \textrm{H}_s \cdot (y_{s,t}^\textrm{cha} + y_{s,t}^\textrm{dis}) \\[\jot] 
    + \lambda^\textrm{PFR}_s \cdot PFR_{s,t} + \lambda^\textrm{EFR}_s \cdot EFR_{s,t} \biggl)   
    \end{array}
    \right\} {\parbox{1cm}{\scriptsize {system costs}}}\\
    \hspace*{8mm}\left. 
    \begin{array}{cc}
    \displaystyle + \sum_{t \in  \mathcal{T}} \sum_{g \in \mathcal{G}}  \left( \psi^{\textrm{max} (y)}_{g,t} + \psi^{\textrm{max} (y^\textrm{st}_g)}_{g,t}  \right.\\
    \hspace*{15mm}\left. + \psi^{\textrm{max} (y^\textrm{sg}_g)}_{g,t} + \psi^{\textrm{max} (y^\textrm{sd}_g)}_{g,t} \right) 
    \end{array}
    \hspace*{12mm}\right\} {\parbox{1cm}{\scriptsize {thermal profits}}}\\ 
    \hspace*{12mm}+
    \left. \sum_{t \in  \mathcal{T}} \sum_{r \in \mathcal{R}} \textrm{CF}_{r,t} \cdot \mathrm{P}_{r}^\mathrm{max} \cdot \psi^{\textrm{CF}}_{r,t} 
    \hspace*{19.5mm}\right\}{\parbox{1cm}{\scriptsize {renewable profits}}} \\
    \left.
    \begin{array}{cc}
    \displaystyle + \sum_{t \in  \mathcal{T}} \sum_{s \in \mathcal{S}} \left(   -\mathrm{E}_{s}^\mathrm{min} \cdot \psi^{\textrm{min}(E_s)}_{s,t}
    + \mathrm{E}_{s}^\mathrm{max} \cdot \psi^{\textrm{max}(E_s)}_{s,t}  \right.\\ [\jot]
    \left. +\psi^{\textrm{max}( y^\textrm{cha}_s)}_{s,t} + \psi^{\textrm{max}(y^\textrm{dis}_s)}_{s,t} + \psi^{\textrm{(dis-cha)}}_{s,t}   \right)   \\ [\jot]
    + 
    \displaystyle \sum_{s \in \mathcal{S}}\left( \mathrm{E}_{s}^\textrm{ini} \cdot  \psi^{\textrm{ini}}_{s,1}  -  \mathrm{E}_{s}^\textrm{end} \cdot \psi^{\textrm{end}}_{s,\textrm{t}^\textrm{end}} \right)
    \end{array}
    \hspace*{1.5mm}\right\}{\parbox{1cm}{\scriptsize {storage profits}}} 
\end{multline}

In the case of energy, the payment is driven by the expression `$\textrm{P}^\textrm{D}_t \cdot \lambda_{t}^{\textrm{E}}$', where it is clear that energy revenues (perceived by energy offerers) should be covered by the trigger, which is the hourly system demand. 

A different case can be seen for the AS payments: the size of the largest credible loss, $\textrm{P}_t^\textrm{Loss}$, triggers the AS market, where the value of $\omega^\textrm{Loss}_t$ represents the incremental change in the AS market with respect to the size of the largest loss. The value of $\omega^\textrm{Loss}_t$ can be computed considering the Karush-Kuhn-Tucker stationarity condition, as shown in eq.~(\ref{eq:price_loss}). It can be demonstrated using eqs.~(\ref{eq:SyncH_price})-(\ref{eq:price_loss}), and considering complementary slackness for frequency-security constraints (\ref{eq:Rocof})-(\ref{eq:qss}), that the expression `$\Omega^\textrm{Loss}_t=\textrm{P}_t^\textrm{Loss} \cdot \omega^\textrm{Loss}_t$' represents the total AS market:
\begin{alignat}{4} \label{eq:AS_market} 
& \Omega^\textrm{Loss}_t=&&\textrm{P}_t^\textrm{Loss}  \cdot \omega^\textrm{Loss}_t   = && \lambda_{t}^{H} \cdot H_t + \lambda_{t}^{PFR} \cdot  PFR_t &&\nonumber\\ 
    & && && + \lambda_{t}^{EFR} \cdot EFR_t \; \;\  \forall t
\end{alignat}

Even though it is clear that the biggest unit triggers the AS market, this market also covers smaller unit outages. Then, as will be discussed in Section \ref{sec:Mechanisms}, making the biggest unit solely responsible for covering the whole of the AS market would not be equitable or consistent. The following section proposes three cost allocation methodologies to address this issue.

\section{Methodologies for Ancillary Services Cost Allocation} \label{sec:Mechanisms}
This section discusses three different methodologies to allocate AS costs to the specific triggers of these services. While these methodologies take as a starting point the AS market defined in Section \ref{sec:Methodology}, they are general enough to be applied to different AS market formulations.

Although electricity consumers directly bear under-frequency containment costs (i.e.,~costs are socialised), we argue that generators should be responsible for covering these costs. Then, the cost allocation is proposed as a way to incentivise behaviour that benefits social welfare, by reflecting the system-integration cost of the different generators. This cost allocation mechanism would shape investment in the future generation mix, enabling the market to guide all players to reduce overall AS cost in low-inertia, RES-based grids.

\begin{figure}
    \centering
    \includegraphics[trim={0.88cm 0 0.2cm 0},clip,width=1.1\linewidth]{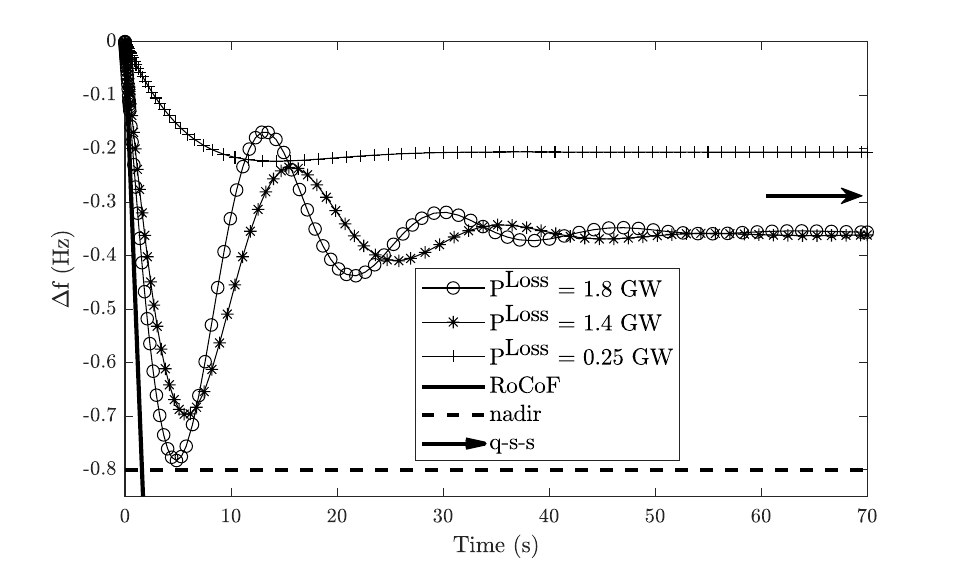}
    \caption{Schematic of system frequency following three different generator contingencies of varying sizes.}
    \label{fig:freq_drop}
    \vspace*{-4mm}
\end{figure}

First, it is important to clarify the relevance of the contingency sizes. In Fig.~\ref{fig:freq_drop}, the system operator would secure the system against the largest loss, which takes a value of $\textrm{P}^\textrm{Loss}=1.8\textrm{GW}$ in GB. This contingency would stress the system to its stability limit, leading to the maximum acceptable frequency deviation. However, the second largest loss, $\textrm{P}^\textrm{Loss}=1.4\textrm{GW}$, would also use the AS that the system operator procured for covering the largest loss. Hence, the second largest unit should be responsible for, at least a portion, of the AS market.

Outages of smaller units can also deviate frequency from its nominal value, as is the case for $\textrm{P}^\textrm{Loss}=0.25\textrm{GW}$ in Fig.~\ref{fig:freq_drop}. However, it is worth analysing whether the necessary AS to cover this small contingency are either (i) only computed by explicit consideration of the frequency-security constraints within the UC; or (ii) are simply a by-product of the energy-only dispatch (e.g., when there is enough inertia and EFR/PFR from the energy market to contain this loss). The former case (i), as was the case for the largest loss, would explicitly need a market for AS. However, for case (ii), if the dispatch provides these services by default, the prices in the AS market would be zero, as an infinitesimal increment of $\textrm{P}^\textrm{Loss}$ would not be binding in the frequency-security constraints. Then, in case (ii) this unit should not be responsible for \emph{any portion} of the AS market. Thus, to perform a cost allocation for these AS, one must consider the impact on the AS market of all possible losses, starting with the largest credible loss, followed by the second largest loss, and smaller losses henceforth. 

To calculate the AS cost allocation for each market player, we compute the AS market that each dispatched unit $i \in \mathcal{N}_t$ would create if it were causing the loss, where $\mathcal{N}_t \subseteq \mathcal{G}  \cup  \mathcal{R} \cup \mathcal{S}$ represents the set of the dispatched units $\{1, 2,..., n-1, n\}$ at hour $t$. 

\begin{figure}
    \includegraphics{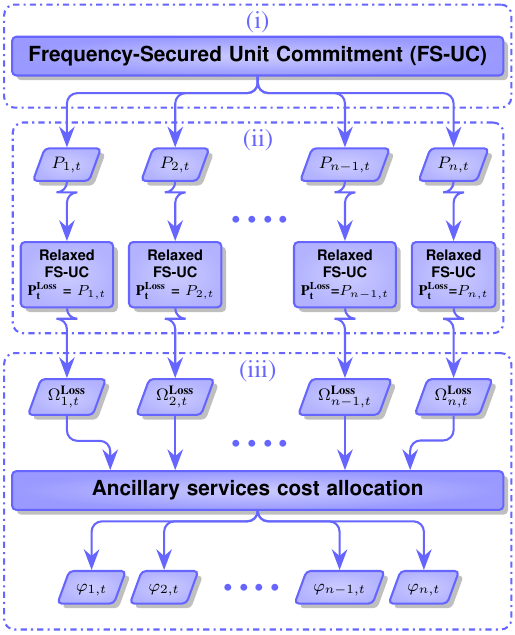}
    \caption{AS cost allocation methodology. Set $\mathcal{N}_t = \{1, 2,..., n-1, n\}$ represents the subset of $\mathcal{G} \cup \mathcal{R} \cup \mathcal{S}$ that is dispatched (discharging in the case of storage units) at hour $t$.}
    \label{FigAS_costAllo}
    \vspace*{-4mm}
\end{figure}

The proposed approach for the cost allocation is summarised in Fig.~\ref{FigAS_costAllo}, where block (i) represents the frequency-secured UC problem described in Sections~\ref{sec:FreqSecUC}, \ref{sec:TotalProvisionAS}, and \ref{sec:FreqConstraints}. In block (ii), a relaxed version of the model is executed for each dispatched unit in $\mathcal{N}_t$, in which parameter $\textrm{P}_t^\textrm{Loss}$ represents the dispatch for that unit determined in block (i). In this stage, we compute the stand-alone AS market expression $\Omega^\textrm{Loss}_{i,t}$ defined in eq.~(\ref{eq:AS_market}), which represents the size of the AS market for each dispatched unit. Finally, in block (iii), the cost allocation methodology for AS is performed. For this task, either of the three methodologies to be presented in Sections~\ref{subsubsec:prop_CA}, \ref{subsubsec:sequential_CA}, and \ref{subsubsec:nucleolus_CA} can be applied.

Without loss of generality, we consider that the set of stand-alone costs $\{ \Omega^\textrm{Loss}_{1,t}, \Omega^\textrm{Loss}_{2,t },...,\Omega^\textrm{Loss}_{n,t }  \}$ is sorted in ascending order, then $\Omega^\textrm{Loss}_{n,t}$ represents the stand-alone AS market for the biggest unit at hour $t$. As stated before, the actual cost of the AS market is associated with the biggest unit in the system. More generally, we consider that the AS market cost function of any coalition of generators has the property that it is equal to the cost of the biggest player in that coalition, i.e., $C(\mathcal{M}_t)=\max\limits_{i \in \mathcal{M}_t}{\{\Omega^\textrm{Loss}_{i,t}\}} \; \forall \mathcal{M}_t \subseteq \mathcal{N}_t$, then the AS market cost allocation represents an \textit{airport problem}.

The \textit{airport problem} \cite{littlechild1977aircraft} is a convex cost allocation game in which $n$ airlines must share the cost of a runaway, which is determined by the largest type of aircraft to land there. This is also the case for AS markets, given that the system operator determines the AS market for the biggest dispatched unit. The fact that the AS market cost allocation is a convex game provides rigorous market guarantees, as shown next. 

\subsection{Proportional AS cost allocation}
\label{subsubsec:prop_CA}
Proportional cost allocation (or `average cost pricing' as named in \cite{moulin1992SerialCSharing}) is one of the simplest ways to allocate costs. It considers that each unit should pay an amount proportional to the stand-alone AS market that it creates, as stated in eq.~(\ref{eq:AS_market}). In this case, the AS cost allocation for each unit is described as follows:
\begin{equation} \label{eq:prop_CA}
    \varphi_{i,t}^{\textrm{Pro}} = \Omega^\textrm{Loss}_{i,t} \cdot
    \frac{ \Omega^\textrm{Loss}_{n,t}  }{\displaystyle  \sum_{i \in \mathcal{N}_t} \Omega^\textrm{Loss}_{i,t}} \;\;\;\;\;   \forall i \in \mathcal{N}_t , \forall t
\end{equation}
Proportional AS cost allocation complies with two main properties \cite{moulin1991axioms}: 

\paragraph{Efficiency} the sum of the cost allocations is the same as the total AS market defined by  $ \Omega^\textrm{Loss}_{n,t} $. Thus, we have:
\begin{equation} \label{eq:prop_CA_eff}
    \displaystyle  \sum_{i \in \mathcal{N}_t} \varphi_{i,t}^{\textrm{Pro}} =    \Omega^\textrm{Loss}_{n,t}  \;\;\;\;\;  \forall t
\end{equation}

\paragraph{Individual rationality} each player pays less than or equal to its stand-alone cost,  $\Omega^\textrm{Loss}_{i,t}$.
\begin{equation} \label{eq:prop_CA_ind_rat}
     \varphi_{i,t}^{\textrm{Pro}} \leq   \Omega^\textrm{Loss}_{i,t}    \;\;\;\;\;  \forall i \in \mathcal{N}_t , \forall t
\end{equation}

However, there are drawbacks to this methodology. Proportional cost allocation does not comply with \textit{coalitional rationality}, i.e., there may be a subset of market players that would be subsidising bigger units. Then, this cost allocation would create cross-subsidies \cite{subsidyFaulhaber1975}. 

\subsection{Shapley value AS cost allocation}
\label{subsubsec:sequential_CA}
Proposed in \cite{shapley1953value}, the Shapley value is a payoff vector, which gives each player their weighted marginal contribution to each coalition. 
Considering a cost function $C(\cdot)$, and $C(\emptyset)=0$, the general formulation for the Shapley value is stated in eq.~(\ref{eq:sh_CA}):
\begin{multline} \label{eq:sh_CA}
    \varphi_{i,t}^{\textrm{SV}} \hspace{-1mm}=\hspace{-2mm} \sum_{\mathclap{\mathcal{M}_t \subseteq \mathcal{N}_t \setminus \{i\}}} \frac{ |\mathcal{M}_t|! (|\mathcal{N}_t|-|\mathcal{M}_t|-1)!}{|\mathcal{N}_t|!} \left[ C(\mathcal{M}_t \cup \{i\}) \hspace{-1mm} - \hspace{-1mm} C(\mathcal{M}_t) \right] \\
    \forall i \in \mathcal{N}_t , \forall t
\end{multline}
Shapley value computation is exponential with the number of players, making its formulation intractable even for a few dozens of players \cite{CREMERS2023120328}. 
However, as the AS market represents an `airport problem', \cite{littlechild1973simple} demonstrated that the Shapley value coincides with the \textit{sequential cost allocation} methodology. The principle behind sequential cost allocation (\textit{serial cost-sharing} in \cite{moulin1994serial}, \textit{sequential equal contribution rule} in \cite{thomson2007cost}, or \textit{standard solution} in \cite{albizuri2015non}) is simple: if two market players create the same AS market, they share the cost equally; if one of them needs a bigger AS market, that player will be responsible for the difference. Following the demonstration in \cite{littlechild1973simple}, we state a compact expression for the Shapley value considering the formulation in \cite{albizuri2015non}:
\begin{equation} \label{eq:seq_CA}
    \varphi_{i,t}^{\textrm{SV}} = \sum_{k=1}^{i} \frac{\Omega^\textrm{Loss}_{k,t} - \Omega^\textrm{Loss}_{k-1,t}}{n+1-k} \;\;\;\;\; \forall i \in \mathcal{N}_t , \forall t
\end{equation}

Thus, for a hypothetical electricity market with two generators $\{1,2\}$, with AS stand-alone markets $\Omega^\textrm{Loss}_{1,t}\leq\Omega^\textrm{Loss}_{2,t}$, the Shapley value AS cost allocation would be: 
\begin{equation} \label{eq:jk2_seq_CA}
   \varphi_{1,t}^{\textrm{SV}} =   \frac{\Omega^\textrm{Loss}_{1,t}}{2} , \;\;\;\;\; 
   \varphi_{2,t}^{\textrm{SV}} =  \Omega^\textrm{Loss}_{2,t}-\frac{\Omega^\textrm{Loss}_{1,t}}{2} \;\;\;\;\; \forall t  \\
\end{equation}

Shapley value complies with several properties, of which we highlight:

\paragraph{Efficiency}\label{prop_eff_SV} as demonstrated in \cite{shapley1953value}, holds.

\paragraph{Coalitional rationality}\label{prop_coalRat_SV} as the `airport problem' is a convex game, the Shapley value lies in the core, therefore coalitional rationality holds \cite{young1985cost}. This implies that coalitions do not pay more than what is required to satisfy their need, meaning that this is an \textit{equitable} cost allocation \cite{hu2006allocation}. This is shown in eq.~(\ref{eq:sh_CA_coalRat}): 
\begin{equation} \label{eq:sh_CA_coalRat}
    \sum_{\mathclap{i \in \mathcal{M}_t}}\varphi_{i,t}^{\textrm{SV}} \leq C(\mathcal{M}_t)  = \max\limits_{i \in \mathcal{M}_t}{\{\Omega^\textrm{Loss}_{i,t}\}} \;\;\;\;\; \forall \mathcal{M}_t\subseteq \mathcal{N}_t, \forall t 
\end{equation}
Note that \textit{individual rationality} is also satisfied by this cost allocation method, given that this is a particular case of coalitional rationality as demonstrated in \cite{shapley1953value}.

According to \cite{subsidyFaulhaber1975}, as this cost allocation satisfies `efficiency' and `coalitional rationality', then it represents a subsidy-free cost allocation.  

\begin{table*}[t]
    \captionsetup{justification=centering, textfont={sc,footnotesize}, labelfont=footnotesize, labelsep=newline}
    \centering
    \renewcommand{\arraystretch}{1.2}
    \caption{Characteristics of power generators, RES and storage}
    \label{tab:GenerationMix}
    \begin{tabular}{p{2.7cm}|p{0.9cm}|p{0.8cm}|p{1.17cm}|p{1.17cm}| p{1.17cm}|p{1.17cm}| p{0.8cm}| p{0.8cm}| p{0.5cm}| p{1.22cm}| p{1.22cm}} %
                                 & Big Nuclear     & Nuclear     & CCGT   & OCGT  & Biomass & BECCS  & Offshore Wind   & Onshore Wind  & Solar PV   & PHES       & BESS\\ \hline
    Installed capacity (GW)       & 1.8            & 3.2        & 25      & 1      & 3  & 1   & 50.4        & 30      & 42      & 4.8     & 20\\ \hline
    Power range (MW) ($\textrm{P}^\textrm{msg}_g, \textrm{P}^\textrm{max}_g$), ($0, \textrm{P}^\textrm{max}_r$), ($-\textrm{P}^\textrm{max}_s, \textrm{P}_{s}^\textrm{msg}, \textrm{P}^\textrm{max}_s$)      &1800, 1800  &1600, 1600  &500,1000  &50,100  &450,500 &450,500 &0,1800    &0,600  &0,250    &-400,0,400    &-100,0,100 \\ \hline
    Inertia constant (s)  &5  &5   &5  &5  &5 &5 &N/A  &N/A    &N/A   &5   &N/A \\ \hline
    Response type    &N/A      &N/A     &PFR   &PFR  &PFR  &PFR &N/A  &N/A  &N/A    &PFR   &EFR\\ \hline
    Response capacity (MW)   &N/A   &N/A  &$30\% \cdot \textrm{P}^\textrm{max}_g$    &$30\% \cdot \textrm{P}^\textrm{max}_g$  &$20\% \cdot \textrm{P}^\textrm{max}_g$  &$20\% \cdot \textrm{P}^\textrm{max}_g$ &N/A        &N/A   &N/A   &$20\% \cdot \textrm{P}^\textrm{max}_s$   &$5\% \cdot \textrm{P}^\textrm{max}_s$\\ \hline
    Energy price offer $\lambda^\textrm{E}_g$, $\lambda^\textrm{E}_r$, $\lambda^\textrm{E}_s$ (\pounds/MWh)  &78    &78  &99  &222   &98  &138 &47  &45  &39   &60  &50\\ \hline
    Inertia price offer $\lambda^\textrm{H}_g$, $\lambda^\textrm{H}_s$ (\pounds/MWs)      &1 &1  &2  &5  &4 &3.5  &N/A  &N/A  &N/A  &1  &N/A\\ \hline
    FR price offer $\lambda^\textrm{PFR}_g$, $\lambda^\textrm{PFR}_s$, $\lambda^\textrm{EFR}_s$ (\pounds/MW)      &N/A &N/A  &3  &7  &4.5 &4.5  &N/A  &N/A  &N/A  &5  &10\\ \hline
    Fuel costs (\pounds/MWh)   &8 &8  &90  &137  &44 &84  &0  &0  &0  &0  &0\\ \hline
    \end{tabular}
\end{table*}

\subsection{Nucleolus AS cost allocation}
\label{subsubsec:nucleolus_CA}
The nucleolus is a payoff vector which makes the coalition that is worst off as well off as possible, then doing the same with the second worst off coalition, and so on; in this way, nucleolus minimises the maximum dissatisfaction for each coalition \cite{schmeidler1969nucleolus}. As proposed in \cite{kopelowitz1967computation}, the nucleolus is calculated as a series of linear programs. Its calculation is challenging as it involves the lexicographical minimisation of $2^n$ excess values, where $n$ is the number of players. Then, the computation of the nucleolus becomes extremely demanding for more than 20 players \cite{benedek2021finding}. 

Nevertheless, \cite{littlechild1974simple} determined a simple form for the nucleolus in the case of the `airport problem', that can be defined inductively. Here we consider the compact formulation exposed in \cite{owen2013game}. Without loss of generality, we consider $m$ technology types of units where $\mathcal{N}_{j,t}$ and $n_{j,t}$ are the set and the number of units of technology type $j$, respectively. Then $\mathcal{N}_t=\bigcup_{j=1}^{m}\mathcal{N}_{j,t}$ and $|\mathcal{N}_t| = \sum_{j=1}^{m}n_{j,t}$. As in the previous case, we consider the ordered stand-alone costs per technology type, thus $\{ \Omega^\textrm{Loss}_{1,t}, \Omega^\textrm{Loss}_{2,t },...,\Omega^\textrm{Loss}_{m,t }  \}$ is sorted in ascending order, and we have $\Omega^\textrm{Loss}_{m,t} = \Omega^\textrm{Loss}_{n,t}$. 
For this case, the objective function for the sequence of linear programs that determine the nucleolus, $\alpha_{q+1,t}$, can be determined inductively as the set of excesses depicted in eqs.~(\ref{eq:excess_1})-(\ref{eq:excess_q}). Where  $ l_{k,t} = \sum_{j=1}^{k}n_{j,t}$,  $ l_{k,t}^r = \sum_{j=k_{r-1}+1}^{k}n_{j,t}$, and $k_{q+1,t}$ is the value of $k$ which gives the minimum $\alpha_{q+1,t}$. This procedure is repeated while $k_q \leq m-2$, then when $k_q=m$ or $k_q=m-1$, the nucleolus is completely determined and shown in eqs.~(\ref{eq:nucl_cluster})-(\ref{eq:nucl_individual}).
\begin{alignat}{3}
    &\alpha_{1,t} && =   -\min \Biggl\{  && \min_{1 \leq k \leq m-1} \biggl\{ \frac{\Omega^\textrm{Loss}_{k,t}}{l_{k,t}+1}\biggl\}, \frac{\Omega^\textrm{Loss}_{m,t}}{l_{m,t}}  \Biggl\}  \quad \forall t \label{eq:excess_1}\\
    &\alpha_{q+1,t} && =   -\min  \Biggl\{  &&    \min_{k_q+1 \leq k \leq m-1} \biggl\{  \frac {\Omega^\textrm{Loss}_{k,t}+\sum_{r=1}^q l_{k_r,t}^r\cdot\alpha_{r,t}}{l_{k,t}^{q+1}+1}    \biggl\},   \nonumber\\
    & && && \frac{\Omega^\textrm{Loss}_{m,t}+\sum_{r=1}^q l_{k_r,t}^r\cdot\alpha_{r,t}}{l_{m,t}^{q+1}}    \Biggl\}   \; \qquad \forall t \label{eq:excess_q}\\
    &\varphi_{j,t}^{\textrm{Nucl}} && = -\alpha_{q,t} && \qquad k_{q-1}+1 \leq j \leq k_q, \; \forall t  \label{eq:nucl_cluster}\\
    &\varphi_{i,t}^{\textrm{Nucl}} && = \; \varphi_{j,t}^{\textrm{Nucl}}  && \qquad i \in \mathcal{N}_{j,t}, \; \forall t \label{eq:nucl_individual}
\end{alignat}

Again, considering two generators with AS stand-alone markets $\Omega^\textrm{Loss}_{1,t}\leq\Omega^\textrm{Loss}_{2,t}$, the nucleolus AS cost allocation would be as shown in eqs.~(\ref{eq:nucl_ex_alpha})-(\ref{eq:nucl_ex_phi}). In this case with $n=2$, the result is the same as for Shapley value (\ref{eq:jk2_seq_CA}), but it would differ for cases with $n\geq3$. 
\begin{alignat}{3} 
    \alpha_{1,t} = -\min \biggl\{ \frac{\Omega^\textrm{Loss}_{1,t}}{2}, \frac{\Omega^\textrm{Loss}_{2,t}}{2}  \biggl\} = -\frac{\Omega^\textrm{Loss}_{1,t}}{2} \qquad \forall t \label{eq:nucl_ex_alpha}\\
   \varphi_{1,t}^{\textrm{Nucl}} =   \frac{\Omega^\textrm{Loss}_{1,t}}{2}, \qquad 
   \varphi_{2,t}^{\textrm{Nucl}} =  \Omega^\textrm{Loss}_{2,t}-\frac{\Omega^\textrm{Loss}_{1,t}}{2} \qquad   \forall t  \label{eq:nucl_ex_phi}
\end{alignat}

Nucleolus complies with several properties, such as:
\paragraph{Efficiency}\label{prope_eff_nucl} nucleolus satisfies \textit{equal treatment of equals}, \textit{last-agent cost additivity}, and \textit{consistency}, as stated in \cite{hwang2012characterization}, therefore it satisfies efficiency.

\paragraph{Coalitional rationality}\label{prope_coalRat_nucl} for convex games, the core is non-empty \cite{shapley1971cores}. Moreover, if the core is non-empty, the nucleolus always exists and lies in the core \cite{benedek2021finding}. Thus, nucleolus is also an \textit{equitable} cost allocation. 

\paragraph{Consistency}\label{prop_stab_Nucl} nucleolus is a \textit{consistent} cost allocation rule because, when fixing one player's allocation, it gives the same solution to the other players \cite{young1985cost}. The property of \textit{consistency} relates to the concept of a valid, acceptable, or fair cost allocation, that should be seen in the same way by any coalition \cite{hu2006allocation}. 

While nucleolus could be computationally demanding for certain games, its computation is straightforward for the AS market, given its structure as an \textit{airport problem}. Therefore, we propose this method for allocating the cost of frequency-containment AS in power grids, as it has one important advantage: unlike Shapley value, it complies with the property of \textit{consistency}, therefore guaranteeing that market players are incentivised to remain in this cooperative game. Nonetheless, in the following section, a comparison of the performance of these three methods is provided, using as a test case a realistic future representation of the GB generation mix.

\section{Case Study} \label{sec:Results}
In this section, we highlight the applicability of the cost allocation methodologies proposed in Section~\ref{sec:Mechanisms}, in combination with the AS market design introduced in Section~\ref{sec:Methodology}. 

This case study is based on the `Leading the Way' scenario within National Grid's Future Energy Scenarios (FES) \cite{NationalGridFES2022}, which provides an ambitious pathway towards decarbonisation in GB. National demand reaches a peak of 62.7 GW, following a typical profile for the GB system considering daily trends. Generation technical parameters are included in Table~\ref{tab:GenerationMix}. Energy price offers are based on the Levelised Cost of Electricity (LCOE) in GB \cite{beis2020electricity}. Inertia price offers are based on no-load plus start-up costs, i.e., the cost of being synchronised. FR price offers are based on National Grid's FR market information \cite{FRReport2022}. Fuel costs are also considered from \cite{beis2020electricity}. Renewable hourly profiles are obtained from \cite{pfenninger2016renewables}. 

Frequency limits are based on GB regulation, considering $\textrm{RoCoF}_\textrm{max}=1\textrm{Hz/s}$ and $\Delta f_\textrm{max}=0.8\textrm{Hz}$. Response services, EFR and PFR, must be fully delivered in $\textrm{T}_\textrm{EFR}=1\textrm{s}$ and $\textrm{T}_\textrm{PFR}=10\textrm{s}$, respectively.

The optimisation model is formulated in YALMIP \cite{YALMIP}, while the solver used for low-level computing is Gurobi \cite{gurobi}.

\subsection{Market design for ancillary services} \label{AS_MD}
Considering the generation mix in Table~\ref{tab:GenerationMix}, the largest loss, i.e.~the value of $\textrm{P}_t^\textrm{Loss}\textrm{=}P_{n,t}$, can be driven by either the Big Nuclear plant or an Offshore wind farm. The frequency-secured UC will schedule the appropriate amount of AS to face any outage, following the \textit{N}{-1} reliability requirement at any given hour.

\begin{figure}[t!]
    \centering
    \includegraphics[width=1\linewidth]{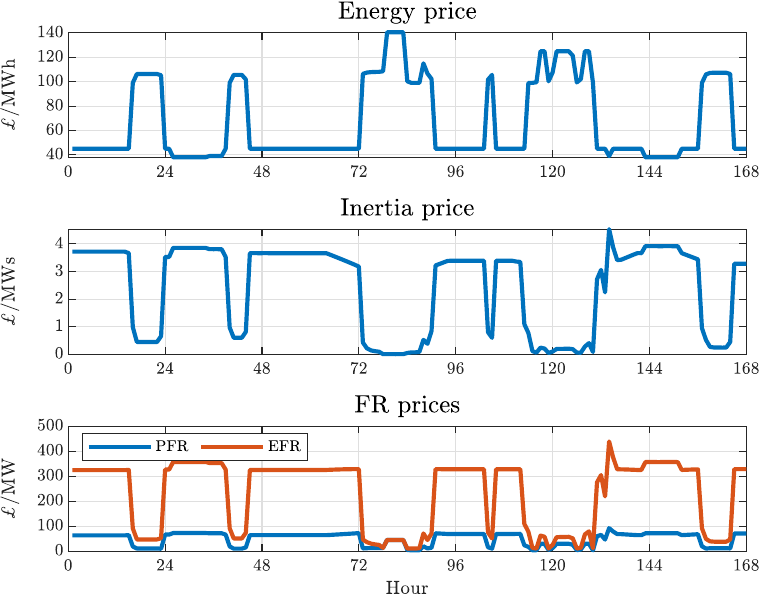}
    \caption{Energy price ($\lambda^{E}_t$) and AS prices ($\lambda_t^{\textrm{H}}, \lambda^{PFR}_t, \lambda^{EFR}_t$).}
    \label{fig:prices}
    \vspace*{-2mm}
\end{figure}

The convex formulation previously discussed, i.e., the relaxed version of the frequency-secured UC, allows the computation of dual variables that define inertia and FR prices, as stated in eqs.~(\ref{eq:SyncH_price})-(\ref{eq:priceEFR}). For a weekly operation, energy and AS prices are shown in Fig.~\ref{fig:prices}. As can be seen, energy and AS prices show an opposite trend: when energy prices are high, AS prices become less relevant. This is because high energy prices are driven by the dispatch of thermal units, which provide inertia as a by-product of energy. Then, even though the model still schedules the necessary amount of AS in these hours, an additional unit of AS becomes practically irrelevant for the optimisation, dropping their prices in this marginal pricing setup. 

\begin{figure}
    \centering
    \includegraphics[trim={0 0.5cm 0 0},clip,width=1\linewidth]{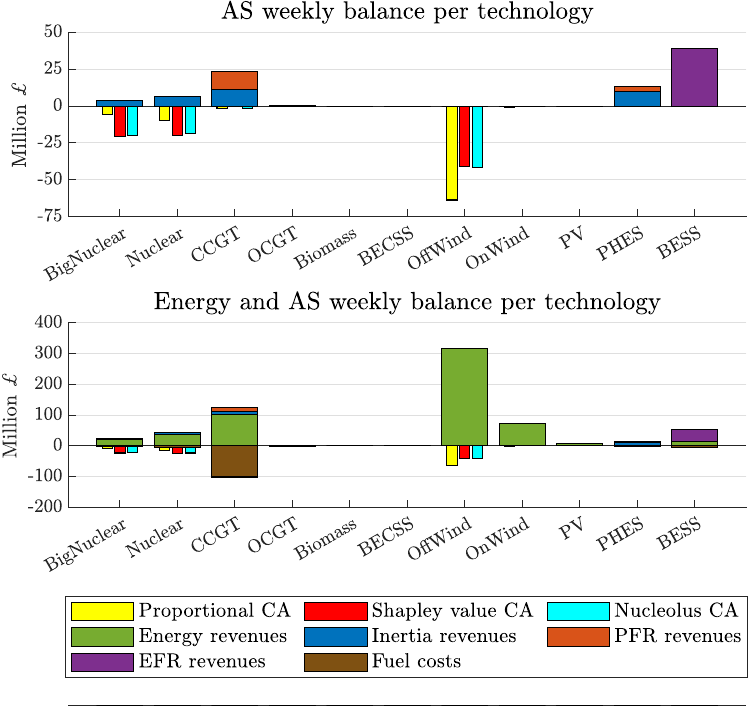}
    \caption{AS market and cost allocation (CA) per technology (on top). Energy and AS markets, fuel costs and cost allocation per technology (bottom).}
    \label{fig:bar2_allCA_balance_EAS_pTech}
    \vspace*{-2mm}
\end{figure}

The summary of weekly revenues is detailed per technology in Fig.~\ref{fig:bar2_allCA_balance_EAS_pTech}. The results show that EFR represents 45\% of the AS market, followed by inertia and PFR with 37\% and 18\%, respectively. Technologies such as BESS, combined-cycle gas turbines (CCGT) and PHES get the largest share of the AS market, with 45\%, 27\%, and 15\% of the total, respectively. 

\begin{figure}
    \centering
    \includegraphics[width=0.95\linewidth]{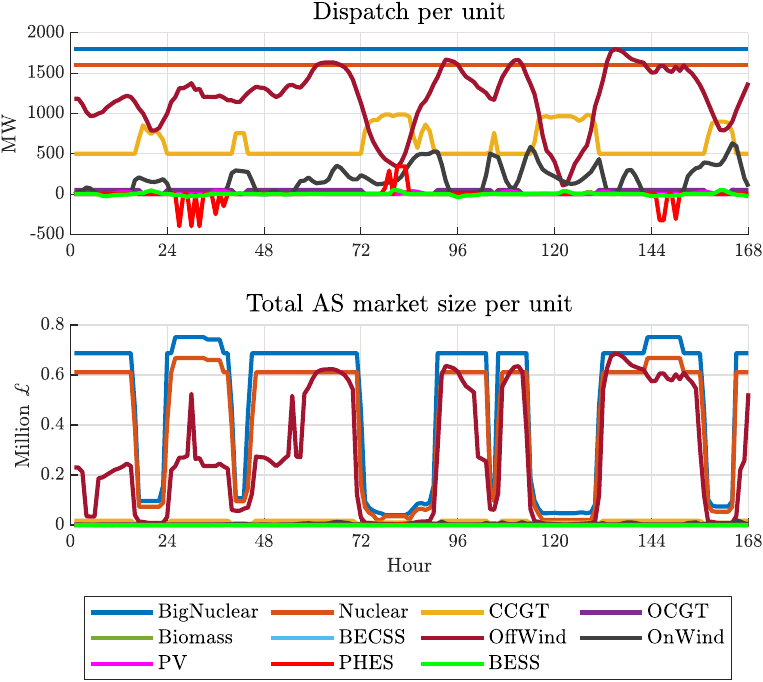}
    \caption{Dispatch ($P_{g,t}, P_{r,t}, P_{s,t}^\textrm{cha}, P_{s,t}^\textrm{dis}$, on top) and AS market size ($\Omega^\textrm{Loss}_t$, bottom) per unit.}
    \label{fig:allRev_disp_marketAS_pGen}
    \vspace*{-4mm}
\end{figure}

The weekly profile of the total AS market, equal to the total AS revenues for all market players, can be seen in Fig.~\ref{fig:allRev_disp_marketAS_pGen} (bottom) for the Big Nuclear plant (in blue), which is the unit that defines the actual size of the AS market for most of the hours. Note that, as was stated in Section \ref{sec:Mechanisms}, most smaller units other than nuclear and offshore wind would not create any AS market whatsoever, given the large difference in rated capacity between these two technologies and the rest.

However, as was shown in Fig.~\ref{FigAS_costAllo}, we compute the size of the AS that each unit would create, i.e. the $\Omega^\textrm{Loss}_t$ term for each unit. These terms are used as input for the cost allocation performed in the next section.

\subsection{Cost allocation of ancillary services} \label{AS_CA} 
The proportional, Shapley value, and nucleolus cost allocations described in Section~\ref{sec:Mechanisms} are applied in this section, in order to compare their performance in a realistic test case. 

\begin{figure}
    \centering
    \includegraphics[trim={0 0.5cm 0 0},clip,width=1\linewidth]{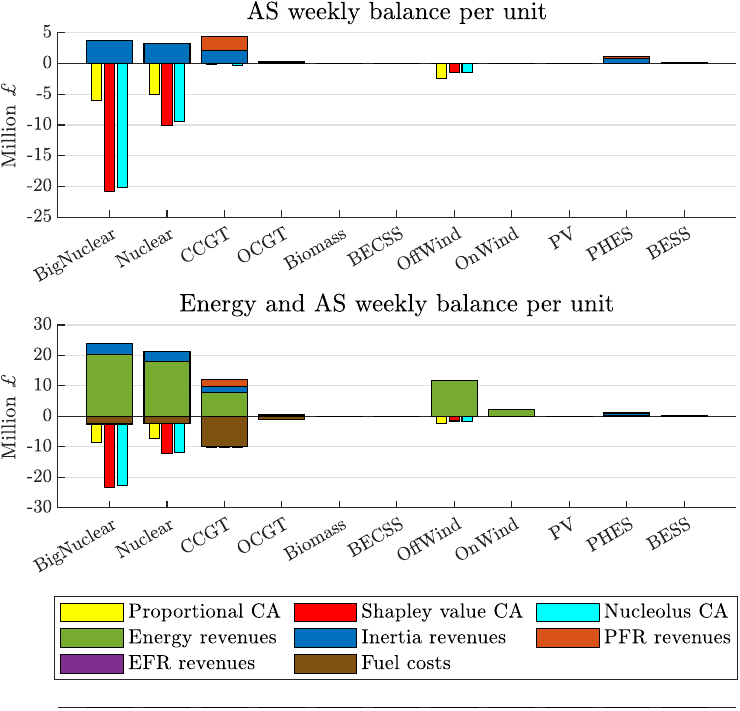}
    \caption{AS market and cost allocation (CA) per unit (on top). Energy and AS markets, fuel costs and cost allocation per unit (bottom).}
    \label{fig:bar2_allCA_balance_EAS_pGen}
    \vspace*{-6mm}
\end{figure}

Fig.~\ref{fig:bar2_allCA_balance_EAS_pGen} shows that Shapley value and nucleolus have minor differences, since both methodologies lie in the core for this convex game, as was stated in Sections~\ref{subsubsec:sequential_CA} and \ref{subsubsec:nucleolus_CA}. As can be seen, the nucleolus implies a slightly smaller cost allocation than the Shapley value for big units such as Big Nuclear (-3\%) and Nuclear (-6\%), since nucleolus makes these units that are worst off as well as possible, as discussed in Section~\ref{subsubsec:nucleolus_CA}. On the other hand, the nucleolus increases the cost for Offshore wind units by 1.5\%, compared with the Shapley value, compensating  the small discount for nuclear plants and therefore balancing out the payments. Even though the Shapley value is an \textit{equitable} cost allocation, while nucleolus is both \textit{equitable} and \textit{consistent}, both methodologies show minor differences for the convex cost function being considered here. 

Fig.~\ref{fig:bar2_allCA_balance_EAS_pGen} also shows that, for the case of the Big Nuclear unit, the proportional cost allocation would assign a cost to this unit $\sim$70\% lower than the one that would be assigned using Shapley value or nucleolus instead. The opposite trend applies to Offshore wind, for which proportional cost allocation assigns a payment $\sim$55\% higher than the one that would be assigned by Shapley value or nucleolus. This creates a cross-subsidy that would be transferred from offshore wind farms to nuclear plants, demonstrating that proportional cost allocation would create undesirable cross-subsidies in the AS market, as was discussed in Section~\ref{subsubsec:prop_CA}. 

The summary of cost allocation results per technology is shown in Fig.~\ref{fig:bar2_allCA_balance_EAS_pTech}. As can be seen, AS cost is mainly distributed among Big Nuclear, Nuclear and Offshore wind generators. In the proportional cost allocation, Big Nuclear (which includes just one unit) and Nuclear (which includes two units) account for 7\% and 12\% of the total market, respectively. With Shapley value, Big Nuclear and Nuclear account for 25\% and 24\%, respectively, amounting to a total of 50\% of the whole cost allocation. Using nucleolus, the Big Nuclear and Nuclear generators account for 24.5\% and 23\%, respectively, accounting both these groups for 48\% of the total cost allocation. 
It is important to highlight that, even though Big Nuclear comprises just one unit, it pays more than two regular-size Nuclear units combined, in both Shapley value and nucleolus cost allocations. This is because Big Nuclear creates the largest AS market, and is the unit that drives the largest possible contingency most of the hours, as seen in Fig.~\ref{fig:allRev_disp_marketAS_pGen}.

Finally, Offshore wind is the technology responsible for more than 77\% of total AS costs using a proportional allocation, while the share of total cost drops to 50\% for the Shapley value and nucleolus allocations. Even though the rated capacity of any single Offshore wind farm is the same to that of the Big Nuclear unit, and the system contains several Offshore wind farms of this size while only one Big Nuclear plant, the Offshore wind fleet is not always dispatched at full capacity (see Fig.~\ref{fig:allRev_disp_marketAS_pGen}), therefore its impact on the AS market is notably less significant than that of the nuclear units.

\vspace*{-0.5mm}

\section{Conclusion and Future Work} \label{sec:Conclusion}
In this paper, a framework for allocating the cost of procuring frequency-containment AS has been proposed and analysed. The goal is to make power system players that cause the need for ancillary services in the first place, accountable for bearing these costs, although in an equitable and consistent manner. 

Three main cost allocation methodologies have been considered, namely proportional, Shapley value, and nucleolus, demonstrating their theoretical properties when applied to the AS market.
It was shown that proportional cost allocation can create cross-subsidies among market players, an undesirable outcome as the cost allocation would not be equitable. On the other hand, both Shapley value and nucleolus are equitable cost allocation mechanisms. Although these two methods are computationally burdensome and potentially intractable for general cooperative games, we demonstrate their full applicability to the AS market for any generic power grid, given its characteristics as an \textit{airport problem}. When applied to the expected future British generation mix, Shapley value and nucleolus show only slight differences. However, nucleolus is the preferred choice for the AS market, given that it meets the critical property of \textit{consistency}, meaning that market players do not have any incentive to leave this cooperative game.

The aim of this work is to provide the electricity market with tools for leading all society to a decarbonised future in an efficient, cost-effective manner. This AS cost allocation methodology would make generators responsible for their system-integration costs. Therefore, when planning future investments, owners would account for these system-integration costs and potentially decide to build smaller plants (which would then drive contingencies of smaller size), or invest in system supporting assets to minimise their `harm' to the grid (e.g., synchronous condensers to increase inertia).

For future work, it would be relevant to fully understand how the proposed cost allocation mechanism would shape future investments. As the proposed mechanism would reduce investment in large units, this would be countered by cost reductions due to economies of scale. Then, it would be essential to understand this balance to determine the key investments for the decarbonisation pathway.

\appendices
\section{Lagrangian Function for AS System-Wide Constraints} \label{ap:Lagrangian_systemwide}
Prices for provision of AS such as inertia (\ref{eq:SyncInertia}), PFR (\ref{eq:pfr_total}), and EFR (\ref{eq:efr_total}) are based on their system requirement, reflected through the frequency-security system-wide constraints RoCof (\ref{eq:Rocof}), nadir (\ref{eq:nadirSOC}), and q-s-s (\ref{eq:qss}). The Lagrangian function for these specific constraints is stated in eq.~(\ref{eq:Lagrangian}) as follows:
\begin{multline} \label{eq:Lagrangian}
    \mathcal{L} \qquad = \qquad \omega^\textrm{Loss}_t \cdot \left(\textrm{P}_t^\textrm{Loss} -  P^\textrm{Loss}_{t} \right)    \\
    +
    \lambda_{t}^{H} \cdot \left [H_t - \sum_{ g \in \mathcal{G} } \textrm{H}_g \cdot \textrm{P}^\textrm{max}_g \cdot y_{g,t}  
    - 
    \sum_{ s \in \mathcal{S} } \textrm{H}_s \cdot \textrm{P}^\textrm{max}_s \cdot (y_{s,t}^\textrm{dis} + y_{s,t}^\textrm{cha})  \right]  \\
    +
    \lambda_{t}^{PFR} \cdot \left(PFR_t - \sum_{ g \in \mathcal{G} } PFR_{g,t}  - \sum_{ s \in \mathcal{S} } PFR_{s,t} \right)  +\\
    \lambda_{t}^{EFR} \cdot \left( EFR_t - \sum_{ s \in \mathcal{S} } EFR_{s,t} \right)
    +    
    \mu^\textrm{RoCoF}_t \cdot \left( \frac{ P^\textrm{Loss}_{t} \cdot f_0}{2 \cdot \mbox{RoCoF}_\textrm{max}} -  H_t   \right) \\
    +
    \mu_{1,t}^\textrm{nadir} \cdot \Biggl(
    \frac{ H_t }{f_0} 
    - \frac{ EFR_t \cdot \textrm{T}_\textrm{EFR} } {4\Delta f_\textrm{max}}  
    - \frac{ PFR_t } { \textrm{T}_\textrm{PFR} }  \Biggr) \\
    + \mu_{2,t}^\textrm{nadir} \cdot \Biggl(
    \frac{  P^\textrm{Loss}_{t} - EFR_t }{ \sqrt{\Delta f_\textrm{max}} }  
    \Biggr)
    \\
    - \mu_{3,t}^\textrm{nadir} \cdot \Biggl(
    \frac{H_t}{f_0} 
    - \frac{ EFR_t \cdot \textrm{T}_\textrm{EFR} } {4\Delta f_\textrm{max}}  
    + \frac{ PFR_t } { \textrm{T}_\textrm{PFR} }  \Biggr)\\
    +
    \mu^\textrm{q-s-s}_t \cdot \left[  P^\textrm{Loss}_{t} - (EFR_t + PFR_t) \right] 
\end{multline}

%
%
%
%



\ifCLASSOPTIONcaptionsoff
  \newpage
\fi



%



\IEEEtriggeratref{32}

\bibliographystyle{IEEEtran} 
\bibliography{Luis_PhD}

\vskip -1.5\baselineskip plus -1fil

\begin{IEEEbiographynophoto}{Carlos Matamala}
(S'22) is Electrical Engineer from the University of Chile (UCH). He worked at the Energy Center UCH, doing research and consultancy to support decision-makers in the energy sector. Currently, he is pursuing a PhD in Electrical Engineering at Imperial College London, U.K. His research interests include operation, planning, and market design for low-carbon energy systems.
\end{IEEEbiographynophoto}

\vskip -1.5\baselineskip plus -1fil

\begin{IEEEbiographynophoto}{Luis Badesa}
(S'14-M'20) received the Ph.D. degree in Electrical Engineering from Imperial College London, U.K. He is currently Associate Professor in Electrical Engineering at the School of Industrial Engineering and Design (ETSIDI), Technical University of Madrid (UPM), Spain. His research focus is on modelling the operation and economics of low-inertia electricity grids, and market design for frequency-containment services.
\end{IEEEbiographynophoto}

\vskip -1.5\baselineskip plus -1fil

\begin{IEEEbiographynophoto}{Rodrigo Moreno}
(M'05) is currently with the Electrical Engineering Department at the University of Chile and Instituto Sistemas Complejos de Ingeniería (ISCI). His research interests involve power system operations and planning under uncertainty, electricity markets, policy and regulation.
\end{IEEEbiographynophoto}

\vskip -1.5\baselineskip plus -1fil

\begin{IEEEbiographynophoto}{Goran Strbac}
(M'95) is Professor and Chair in Electrical Energy Systems at Imperial College London, U.K. His current research is focused on the optimisation of operation and investment of low-carbon energy systems, energy infrastructure reliability and future energy markets. 
\end{IEEEbiographynophoto}

\end{document}